\documentclass[aps,10pt,pra,twocolumn,amssymb,amsmath,amsfonts,showpacs,floatfix,citeautoscript,preprintnumbers,nofootinbib,superscriptaddress,a4paper]{revtex4-1}
\usepackage{graphicx,times}
\usepackage[latin1]{inputenc}

\def\force{F}
\def\energy{\varepsilon}
\def\wband{\Delta}
\def\door{\Pi}
\newcommand{\inter}[2]{\left\langle{#1}\vert{#2}\right\rangle}
\newcommand{\elem}[3]{\langle{#1}\vert{#2}\vert{#3}\rangle}
\newcommand{\ket}[1]{\left\vert{#1}\right\rangle}
\newcommand{\md}[1]{\vert{#1}\vert}
\def\fidelity{\mathcal{F}}

\begin{document}

\title{Entangling many-body bound states with propagative modes in Bose-Hubbard systems}

\author{Mario Collura}
\affiliation{Institut Jean Lamour, dpt. P2M, Groupe de Physique Statistique, Nancy-Universit\'e CNRS, B.P. 70239, F-54506 Vandoeuvre les Nancy Cedex, France}
\author{Helge Aufderheide}
\affiliation{Department Biological Physics, Max-Planck-Institute for the Physics of Complex Systems, Noethnitzer Strasse 38, 01187 Dresden, Germany}
\author{Guillaume Roux}
\email{guillaume.roux@u-psud.fr}
\affiliation{Laboratoire de Physique Th\'eorique et Mod\`eles statistiques, Universit\'e Paris-Sud, CNRS, UMR8626, 91405 Orsay, France}
\author{Dragi Karevski}
\email{dragi.karevski@ijl.nancy-universite.fr}
\affiliation{Institut Jean Lamour, dpt. P2M, Groupe de Physique Statistique, Nancy-Universit\'e CNRS, B.P. 70239, F-54506 Vandoeuvre les Nancy Cedex, France}

\begin{abstract}
  The quantum evolution of a cloud of bosons initially localized on
  part of a one dimensional optical lattice and suddenly subjected to
  a linear ramp is studied, realizing a quantum analog of the
  ``Galileo ramp'' experiment. The main remarkable effects of this
  realistic setup are revealed using analytical and numerical
  methods. Only part of the particles are ejected for a high enough
  ramp, while the others remain self-trapped. Then, the trapped
  density profile displays rich dynamics with Josephson-like
  oscillations around a plateau. This setup, by coupling bound states
  to propagative modes, creates two diverging condensates for which
  the entanglement is computed and related to the equilibrium
  one. Further, we address the role of integrability on the
  entanglement and on the damping and thermalization of simple
  observables.
\end{abstract}


\maketitle

The last decade breakthrough experiments on ultra-cold atomic gases
have revived the field of strongly correlated
many-body quantum systems~\cite{Bloch2008}, especially within non-equilibrium
aspects~\cite{Dziarmaga2010}.  The very low dissipation rate and long
time phase coherence of these systems allow the investigation of
genuine quantum effects, the role of integrability on thermalization
properties, and the possibility to engineer desired states.  Two
examples are particularly relevant for this study.  The first is Bloch
oscillations (BO)~\cite{Bloch1928} which occur when a particle travels
on a lattice and experiences a constant external force $\force$
(potential ramp): Its momentum drifts with time according to
$q(t)=q(0)+\force t$ (we set $\hbar$ and the lattice spacing equal to
one), but modulo the Brillouin zone, setting the BO period $\tau_{B} =
2\pi/|{\force}|$.  They have been observed in many physical domains :
semiconductors~\cite{Waschke1993}, thermal gases~\cite{BenDahan1996},
photonics~\cite{Agarwal2004}, cold atoms~\cite{Morsch2001}, and
phonons~\cite{He2007}. BO can survive to the many-body regime, with a
damping possibly related to the integrability of the
model~\cite{Kolovsky2003,Gustavsson2008}. The second example is the
release of atoms from a trap, a standard protocol with cold atoms.
Yet, keeping the optical lattice on during the expansion allows one to
handle metastable states~\cite{Heidrich-Meisner2009}, or study
transport phenomena in disordered potentials~\cite{Billy2008}. We
lastly stress that ultra-cold atoms, beyond their fundamental
interest, could be used as devices to mimic optics
experiments~\cite{Bloch2001} or semi-conductor
physics~\cite{Seaman2007}.

\begin{figure}[b]
\includegraphics[width=\columnwidth,clip]{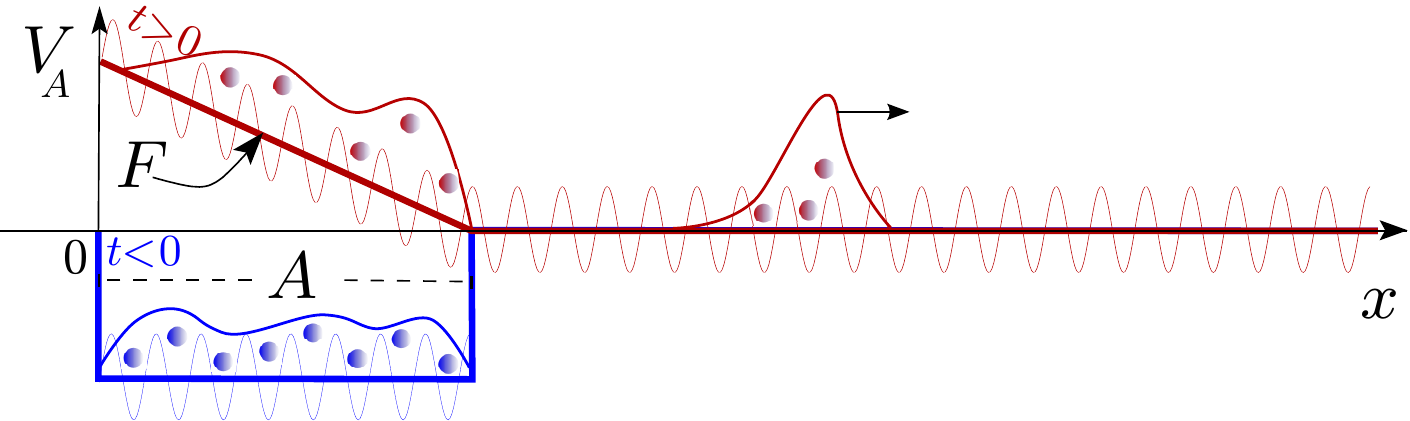}
\caption{(color online) Sketch and notations of the considered setup. \label{fig:model}}
\end{figure}

In this paper, we put forward an experimentally realistic setup which
ejects interacting bosons living on an optical lattice using a linear
potential (see Fig.~\ref{fig:model}). This ``Galileo ramp'' experiment
displays remarkable features that are understood with analytical and
numerical calculations. Only part of the particles is ejected and
forms a wave-packet which shape and number of particles are
determined. Thanks to the initial correlations, these traveling
particles remain strongly entangled with the ones remaining
self-trapped in the initial region, hence creating two diverging and
entangled many-body condensates. In the self-trapping region,
particles exhibit Josephson-like oscillations reminiscent of BO, with
a density plateau due to many-body interferences, and damping
controlled by the integrability of the model. The setup is
particularly versatile, in comparison to analog proposals to create
wave-packets~\cite{Rigol2004}, and relies on the general idea of
coupling bound states to the propagative modes of a lattice.  Results
in the hard-core bosons limit provides a quantitative and intuitive
understanding which allows for straightforward generalizations
(mirrors, beam-splitters,\dots).

\section{Setup description}

\subsection{Hamiltonian}

We describe the setup by means of the one-dimensional (1D) Bose-Hubbard model (BHM) 
\begin{equation*}
\mathcal{H}=-J\sum_j [b^{\dagger}_{j+1}b_j+\text{h.c.}]+\frac{U}{2}\sum_j n_j(n_j-1)+\sum_j V_j(t) n_j
\label{eq:hamil}
\end{equation*}
where $b_j^{\dagger}$ is the bosonic creation operator at site $j$,
$n_j=b_j^{\dagger}b_j$ is the density operator.  $J$ and $U$ are
respectively the hopping and interaction magnitudes while $V_j(t)$ is
the time-dependent external potential.  At times $t<0$, the potential
is a deep box of width $A$ which confines all $N$ particles in region
$\mathcal{A}$, with the average density $\rho=N/A$.  At times $t\geq
0$ (quench), the potential is suddenly changed to a linear ramp $V_j =
V_A-{\force}j$ in region $\mathcal{A}$ and zero elsewhere (see
Fig.~\ref{fig:model}). We provide in Appendix~\ref{app:tunneling} the
criteria to neglect tunneling toward the upper Bloch band.

\begin{figure*}[t]
\centering
\includegraphics[width=1.02\textwidth]{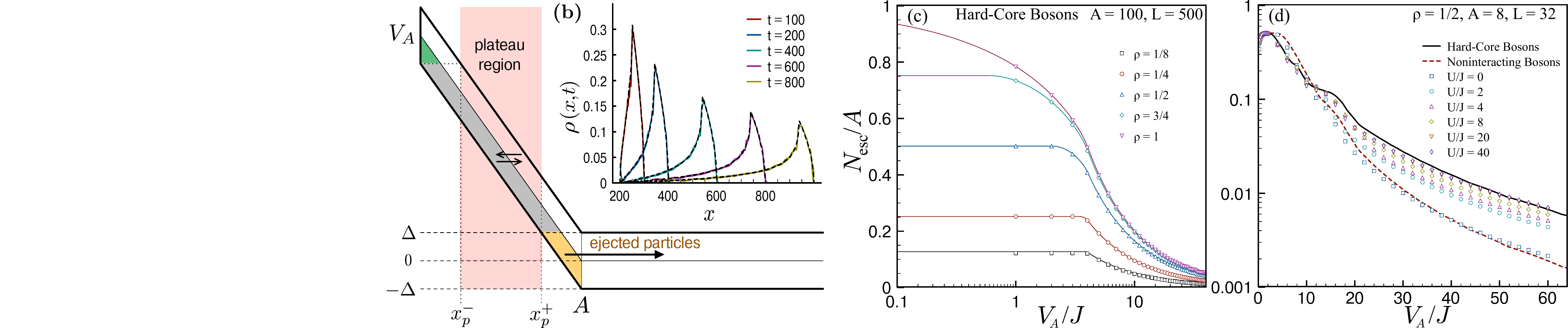}
\caption{(color online) 
\textbf{(a)} time-evolution of the density profile at half-filling for increasing ramp height $V_A=1, 3, 5, 8\;[J]$.  
\textbf{(b)} sketch explaining the hydrodynamic calculations. \emph{Inset}: numerical escaping profiles (full lines) compared to the hydrodynamic approach (dashed lines).
\textbf{(c)} number of escaping HCB particles vs. $V_A$ for various initial densities $\rho$ (hydrodynamic predictions in full lines).
\textbf{(d)} same number, varying the interaction strengths $U/J$ at half-filling, and compared to analytical limiting cases. 
\label{fig:ejected}}
\end{figure*}

For simulations, the setup is embedded in a larger box of size $L$,
and the gas is assumed to be isolated.  Its dynamics is studied via
analytical methods in the non-interacting and Hard-Core Boson (HCB)
$U=\infty$ limits.  HCB dynamics can be solved exactly using the
mapping to a fermion Hamiltonian and diagonalizing the single-particle
physics (see below). Observables are then expressed as determinants
that are computed numerically. The out-of-equilibrium calculations are
based on Refs.~\onlinecite{HCB,Rigol2006}. Otherwise, two ab-initio
techniques are used to compute observables: Lanczos diagonalization
and time-dependent density-matrix renormalization group
(tDMRG)~\cite{DMRG, Vidal,White2004,Daley2004} (see also
Refs.~\onlinecite{Dechiara2008} for a good introduction), which is
particularly useful to compute the entanglement entropy of
Fig.~\ref{fig:entanglement}. Lanczos diagonalization are used with a
cutoff $M$ in the onsite boson number. $M=N$ is taken on the box
configuration (except for $L=18$ for which $M=7$). For $L=32$, the
64bits limitation imposes $M=3<N$. This can participate in the little
difference between the free bosons analytics and the numerics in
Fig.~\ref{fig:ejected}(d).
tDMRG is performed with $M=2$. After sweeping through the chain to
apply the evolution operator, we keep the maximal value of the block
entanglement entropy $S_{max}(t)$ and we update the number of kept
states $m$ at time $t+dt$ to $m(t+dt)= \mathrm{int}[m(t)\times \exp
\Delta S_{max}(t)]$, in order to account for the eventual growth of
entanglement. Typically, starting the evolution with $m(0)\in[16,24]$,
we finish with $m(T)\in[24,110]$ depending on the value of $U$ and
$F$. Comparisons with Lanczos calculations ensured that this is
sufficient to monitor errors for the considered parameters. For these
ab-initio methods, the Hilbert sizes at play are not challenging and
the numerical errors are under control.

\subsection{Hydrodynamic description of HCB} 

In the limit $N,A\gg 1$, after a Jordan-Wigner transformation, a chain
of HCB maps to the free fermion Hamiltonian $\mathcal{H} = \sum_q
\energy_{q} \eta^{\dagger}_q\eta_q$ with the fermionic operators
$\eta_q$ of momentum $q\in[-\pi,\pi]$. The \emph{kinetic} part of the
energy, $-\wband\cos{q}$ with $\wband=2J$, builds up a 1D band of
width $2\wband$ and gives a velocity $v(q) = \wband\sin q$.  The
ground-state is the corresponding Fermi sea of Fermi momentum
$q_F=\pi\rho$. This is a good description of the initial state
provided the well is large and deep enough.  The features of the
density dynamics are well captured by the hydrodynamic limit using the
continuous position variable $x$.  The initial state has approximately
a coarse grained phase-space density $w(x,q) = \frac{1}{2\pi}
\door_{[0,A]}(x)\door_{[-q_F,q_F]}(q)$, where $\door_{[a,b]}$ is the
characteristic function over the interval $[a,b]$.  After the sudden
quench, the energy of a particle at position $x$ is shifted by the
additional potential value $V(x)$.  The dynamics being unitary,
quasi-particles are emitted to the right and to the left on a
trajectory of constant energy $\energy$.  The density of particles
$\rho(x)$ is the sum of the densities of these right and left movers,
to which are attributed the local velocities $v(\pm q(x))$, where
$q(x) = \arccos[(V(x)-\energy)/\wband]$.  This approach allows us to
reconstruct the evolving density profile $\rho(x;t)$.

\section{Wave-packet emission of the Galileo ramp}

\subsection{Wave-packet density profile}

The typical evolution of the density of
HCB after the quench for increasing forces is given on
Fig.~\ref{fig:ejected}(a).  Only part of the bosons are ejected, with
maximum velocity $\wband$, and the wave-packet spreads with
time.  In the HCB hydrodynamic description, the population with energy
$\energy\in[-\wband,\wband]$ is connected to the propagating states
outside $\mathcal{A}$ (see Fig.~\ref{fig:ejected}(b)).  These
particles escape from region $\mathcal{A}$ and propagate toward
$\infty$.  As for BO, the momentum of a right(left) mover of energy
$\energy$ obeys $q^{\pm}(t)=\pm q+Ft$.  From energy conservation, it
converts all its potential energy into kinetic energy when it reaches
the position $A$ at a time $t^{\pm}$, given by ${\force}t^{\pm} = \mp
q + \arccos(-\energy/\wband)$.  After escaping from $\mathcal{A}$,
both travels at constant velocity, giving the ballistic trajectory
$x^{\pm}(t;q,\energy) = \sqrt{\wband^2-\energy^2}(t-t^{\pm})+A$.  At
times larger than $\tau_B$, these particles have all left region
$\mathcal{A}$ and their density profile is the sum of the right and
left movers contributions
\begin{equation}
\rho_{\text{esc}}^{\pm}(x;t)=\int_{-\wband}^{\wband}\!\frac{d\energy}{\force}\int_{Q(\energy)}\!\frac{dq}{2\pi} \delta(x-x^{\pm}(t;q,\energy))\;,
\label{densityesc}
\end{equation}
in which $Q(\energy)$ refers to the yellow domain of
Fig.~\ref{fig:ejected}(b). Fig.~\ref{fig:ejected}(b) shows the very
good agreement, up to small quantum interference effects, between
exact diagonalization data and the hydrodynamic
prediction~\eqref{densityesc}.

\subsection{Number of ejected particles}

The total number $N_{\text{esc}}^{\text{HCB}}$ of escaping HCB is
readily obtained by integrating the initial density on the same domain
\begin{equation*}
\rho(x\leq A,\energy;t=0)=\int_{Q(\energy)} \frac{dq}{\pi}\; \delta(\energy-V(x)+\wband\cos{q})\;.
\label{density0}
\end{equation*}
One obtains that, for a small positive force $\force \leq
\wband(1+\cos{q_F})/A$, all particles are connected with propagative
states, i.e. $N_{\text{esc}}^{\text{HCB}}=N$.  When increasing
$\force$ beyond, some particles remain trapped in region $\mathcal{A}$
and the explicit calculation yields:
\begin{equation}
N^{\text{HCB}}_{\text{esc}} = \frac{\wband}{\pi\force}\big[q_F+\sin{q_F})-g(FA/\Delta)\big]
\end{equation}
with $g(x) =\sqrt{x(2-x)}-(x-1)\arccos(x-1)$ when $\wband(1+\cos
q_F)/A \leq F \leq 2\wband/A$, and $g(x)=0$ when $F \geq 2\wband/A$.
A comparison with simulations is displayed on
Fig.~\ref{fig:ejected}(c), demonstrating the accuracy of the approach.

Lowering interactions naturally modifies $N_{\text{esc}}$. Taking the
opposite case of free bosons, this single-particle physics is solved
along the same lines : 
free bosons are initially gathered in the ground-state
wave-function of the deep well so that the initial
density profile in region $0\leq x\leq A$ is
\begin{equation}
\rho(x;0) = N\frac 2 A \sin^2\left(\frac{\pi x}{A}\right)\;.
\end{equation}
In this regime, all bosons share the same dynamics and the density
profiles of the trapped and outgoing wave-packets depend only
trivially on $N$ which enters as a prefactor.  The initial state thus
consists of $N$ bosons of initial momentum $q_0\simeq \pi/A$.  The
particles leaving the confinement region are again those with energy
$\energy \in [-\wband,\wband]$, corresponding to the space interval
$\vartheta = [\max(0,A-2\wband/\force),A]$. Then, one obtains
$N_{\text{esc}}^{\text{free}} = N$ for $0 \leq F \leq 2\wband/A$
and
\begin{equation}
N^{\text{free}}_{\text{esc}} = 
N \left[\frac{2\wband}{\force A} - \frac{1}{2\pi}\sin\left(2\pi\frac{2\wband}{\force A}\right)\right]\;.
\label{eq:nescU0}
\end{equation}
for $\force \geq 2\wband/A$.  This time, the strong force
scaling is $\force^{-3}$. This last result is also valid for $N=1$, in
which case the single-particle wave-function is in a superposition of
bound and diffusive states. In Fig.~\ref{fig:ejected}(d), numerical
simulations for various $U$ shows how results interpolate between
these two limiting cases with an interaction-dependent behavior at
large $\force$. Note that, surprinsingly, interactions do not favor
escaping at small forces.

\section{Dynamics of trapped particles}

\subsection{Density profile}

As seen from Fig.~\ref{fig:ejected}(a), the trapped HCB exhibit a
rather rich physics with oscillations reminiscent of BO.  This can be
understood in the hydrodynamic approach: We consider the main
contribution coming from trapped particles with energies
$\energy\in[\wband,V_A-\wband]$ (grey area in
Fig.~\ref{fig:ejected}(b)). Neglecting tunneling escapes at the band
edges, the corresponding densities of right/left movers are
\begin{equation*}
\rho_{\text{coh}}^\pm(x;t)=\int_0^{q_F}\!\!\frac{dq}{2\pi}\int_{\wband}^{V_A-\wband}\hspace{-8mm}
d\energy\,\delta\left(\energy-V(x)+\wband\cos(q\pm \force t)\right)\,.
\label{densitypm}
\end{equation*}
These are clearly periodic functions of time, oscillating with period
$\tau_B$.  By introducing $\tilde{q} = q\pm Ft$ in each term and
integrating over energies, the coherent part of the density reads
\begin{equation}
\label{eq:densityt}
\rho_\text{coh}(x;t)=\int_{\force t-q_F}^{\force t+q_F}\frac{d\tilde{q}}{2\pi}
\;\door_{[\wband,V_A-\wband]}(V(x)-\wband\cos{\tilde{q}})\;.
\end{equation}
From~\eqref{eq:densityt}, we see that oscillations are present
provided $\rho<1$ (superfluid regime of HCB). Indeed, at unit filling
$\rho=1$ (Mott state) $q_F=\pi$ and the integration range then covers
a full period, leading to a static trapped density profile.  The
density profile can actually be computed at any time $t$ from
\eqref{eq:densityt} (see below), and a comparison with numerics is
given in Fig.~\ref{fig:oscillations}(a).  We first observe that
incoherent contributions due to high energy modes of energy
$\energy\in[ V_A-\wband,V_A+\wband]$ (green area in
Fig.~\ref{fig:ejected}(b)) not taken into account in
\eqref{eq:densityt} increase both the average and the fluctuations of
the particle density on the left side.  Indeed, due to the free
boundary at $x=0$, high energy particles sharing initially the same
momentum $q$ are not reflected at $x=0$ at the same
time. Consequently, a dephasing appears between them and their
contribution to the total spatial density is somehow incoherent. A
simple way to suppress this incoherent effect is to eject these
particles into a propagative band on the left ($x<0$), as shown in
Fig.~\ref{fig:oscillations}(b) where we see a very good agreement, up
to small interference effects, between the exact numerical results and
the hydrodynamic predictions when the incoherent particles are
removed.

\begin{figure*}[t]
\centering
\includegraphics[width=\textwidth,clip]{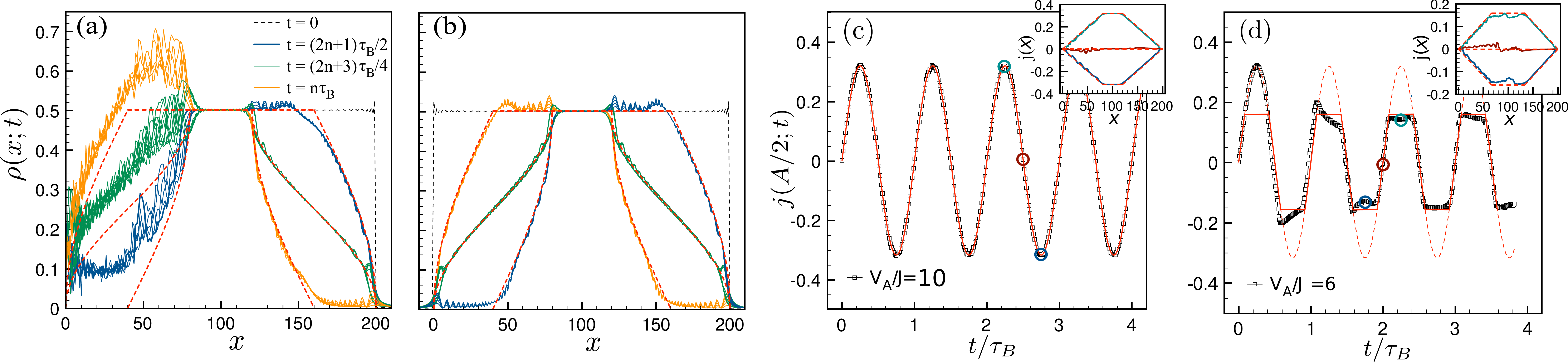}
\caption{(color online) Density profile without \textbf{(a)} and with
  \textbf{(b)} a propagating band opened at the left of $x=0$ at
  different times for $V_A/J=10$, $A=200$ at half-filling $\rho=1/2$
  (hydrodynamic predictions are given by the dashed red lines).
  \textbf{(c)} Corresponding current density at $x=A/2$. 
Inset: Current profiles at
  different times (colors refer to the circles). Full lines are the
  exact results while dashed red lines are the hydrodynamic
  predictions. \textbf{(d)} same but for a smaller ramp $V_A/J=6$.
The straight red line is the expected
  truncated sine function while the dashed red one is the pure sine
  oscillation (see the main text). 
  \label{fig:oscillations}}
\end{figure*}

Another striking feature of Fig.~\ref{fig:oscillations}(a)-(b) is the
existence of a stationary density plateau: The argument of the $\door$
function in \eqref{eq:densityt} reaches its edges for positions
$x_{\text{inf/sup}}$ given by $V(x_{\text{inf}})-\wband\cos \tilde{q}
= V_A-\wband$ and $V(x_{\text{sup}}) - \wband\cos\tilde{q}=\wband$.
For $\max{x_{\text{inf}}}\leq x \leq \min{x_{\text{sup}}}$, the
integral is independent of time and the initial density plateau
survives. Therefore, the condition to have this plateau is
$F>4\wband/A$ and its width is $A-4\wband/F$. The boundaries of the
plateau are given by $x_p^-=\max{x_{\text{inf}}}$ and
$x_p^+=\min{x_{\text{sup}}}$, which can be written
\begin{equation*}
x_p^{\pm} = \frac A 2 \pm\left(\frac A 2-\frac{2\wband}{\force}\right)\;.
\end{equation*}
At each sides of the plateau, an excess density oscillates, being at
the left(right) for (half-)integer multiples of $\tau_B$.

The explicit density profile is easily computed at any time $t$ from
Eq.~\eqref{eq:densityt}.  For example, at integer multiples of the BO
period $\tau_B$, the distribution of the trapped bosons is shifted
maximally to the left (down to $x_{\text{inf}}(q_F)<x_p^-$) and is
given by
\begin{widetext}
\begin{equation}
\rho_{\text{coh}}(x;n\tau_B)=
\begin{cases}
\frac{1}{\pi} \arccos[(V(x)-V_A)/\wband+1]& 0\le x\le x_{\text{inf}}(q_F)\\
\rho&  x_{\text{inf}}(q_F)\le x \le x^+_{p}\\
\rho-\frac{1}{\pi}\arccos[V(x)/\wband-1]&  x_p^+\le x \le x_{\text{sup}}(q_F)\\
0& x_{\text{sup}}(q_F)\le x\le A
\end{cases}
\end{equation}
\end{widetext}
The distribution then propagates from this macroscopic left state to
the macroscopic right one (reached at half periods times
$t=(n+1/2)\tau_B$ with integer $n$). The spatial profile of the right
state is simply deduced from the left one by the transformation
$x\rightarrow A-x$, using the mirror symmetry at $A/2$. These are the
results plotted in Fig.~\ref{fig:oscillations}(a-b).

\subsection{Collective Josephson-like oscillations}

This remarkable pendulum motion of the density is naturally associated
with a flow of particles through the plateau region.  The flow of
particles from the left to the right of the trapping zone and vice
versa gives rise to a periodic current density $j(x;t)$ which, in the
hydrodynamic limit, is simply the sum over all momenta of
$\rho(q)v(q)$, i.e. the quasi-particle currents
$[\rho^+(q)-\rho^-(q)]\wband\sin q$.  In the trapped region the
current density is given by the integral expression
\begin{equation}
j(x;t)={\wband}\!\!\!\!\int\limits_{\force t-q_F}^{\force t+q_F}\!\!\frac{d\tilde{q}}{2\pi}\; \sin \tilde{q} 
\;\door_{[\wband,V_A-\wband]}(V(x)-\wband\cos{\tilde{q}})\;.
\label{eq:currentt}
\end{equation}
For $F>4\Delta/A$, in the plateau region $\Omega$ one has simply from
\eqref{eq:currentt} a spatially constant current (which was expected
from the continuity equation since the local density is constant in
time)
\begin{equation}
j(x;t)=\frac{\wband}{\pi} \sin(q_F)\sin(\force t)\;, 
\end{equation}
which oscillates harmonically with period $\tau_B$. The maximum
current amplitude is obtained at half filling when $q_F=\pi/2$, while
the current is exactly zero in the Mott phase (unitary filling giving
$q_F=\pi$) reflecting the stationarity in time of the trapped density
in that case.  These Josephson-like oscillations are reported in
Fig.~\ref{fig:oscillations}(c) and we see that the hydrodynamic
description matches perfectly the numerical data.  Outside the
plateau, the current density has a spatial variation which is simply
deduced from its integral representation \eqref{eq:currentt}. Again,
as seen in the inset of Fig.~\ref{fig:oscillations}(c), the
hydrodynamic description is very good.

At lower forces, for $2\Delta/A<F<4\Delta/A$, when there is no more a
plateau region but still self-trapped particles, the temporal
evolution of the current is no more given by a pure sine function due
to the interplay between the integration interval and the support of
the door function entering into \eqref{eq:currentt}. For example, at
half filling $q_{F}=\pi/2$ and taking $V_{A}=3\Delta$, one has from
\eqref{eq:currentt} in the middle of the ramp
\begin{equation*}
j(\frac{A}{2},t)
= \left\{\begin{array}{lcl}
\frac{\Delta}{\pi} \sin (\force t)&\text{for} & |\sin(\force t)| \le (V_A/2\Delta-1)\\ 
\Delta/2\pi               &\text{for} & |\sin(\force t)| \ge (V_A/2\Delta-1)  
\end{array}\right. .
\end{equation*}
This truncated sine essentially reflects the escape of particles which
were initially located in the middle of the condensate. Indeed, for
$V_{A}<4\Delta$ the escape locus $x^+_p$ extends over the middle of
the ramp, since $x^{+}_{p}<A/2$, leading to a lowering of the current
intensity after the corresponding particles have been ejected. As a
support of this, notice on Fig.~\ref{fig:oscillations}(d) how during the
first half period, as the particles are moving from the left to the
right and have not yet abandoned the condensate, the current shows a
perfect sinusoidal signal which gets truncated as time goes on.
Notice that, despite a behavior similar to the Josephson effect, the
oscillations stem from a fundamentally different effect -- many-body
interferences with strong interactions -- and without a tunneling
barrier -- the plateau develops \emph{within} the system.

\section{Entanglement entropy between the wave-packets}

\begin{figure}[t]
\includegraphics[width=0.8\columnwidth]{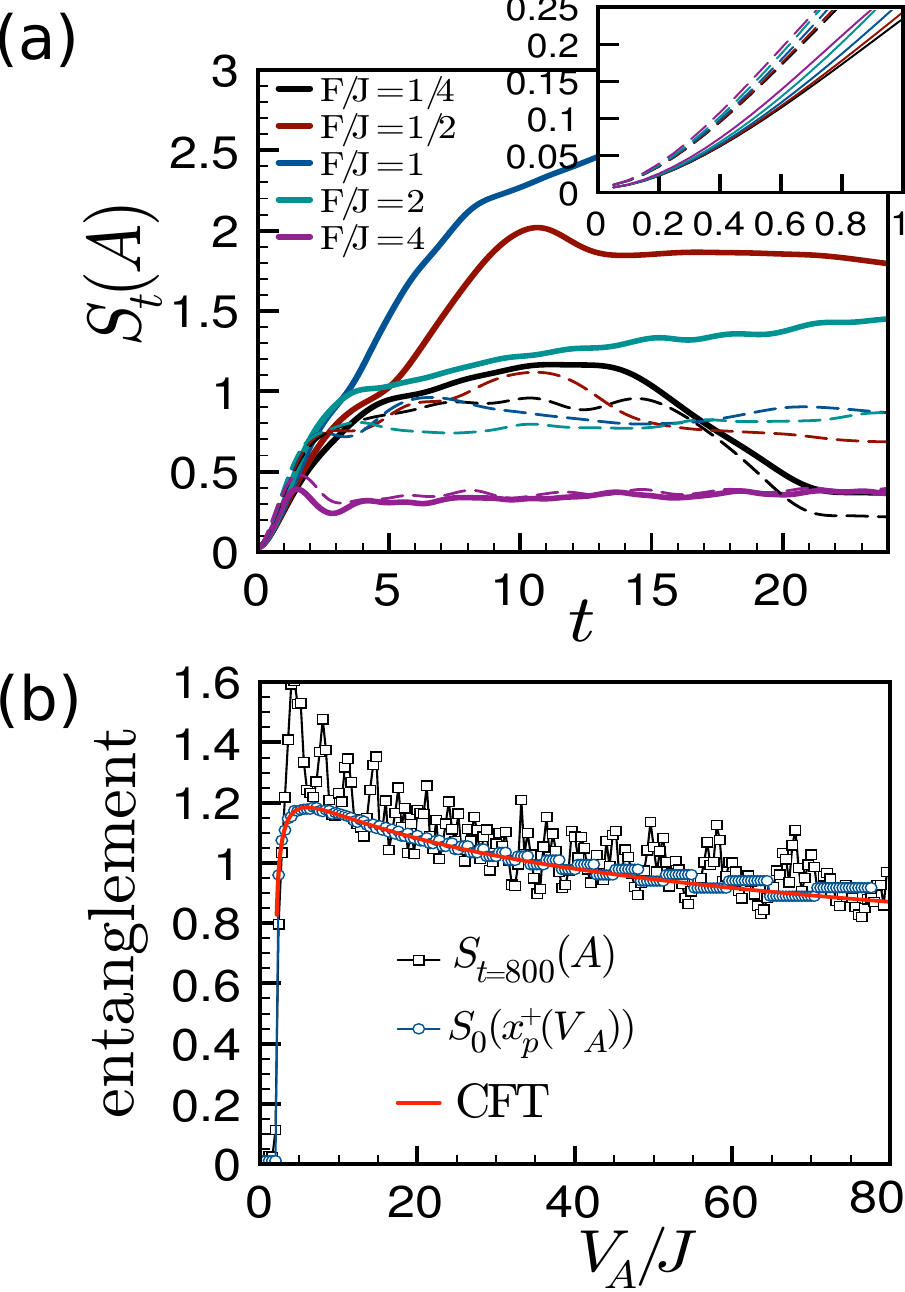}
\caption{\label{fig:entanglement} (color online) \textbf{(a)}
  evolution of the entanglement entropy $S_t(A)$ for an increasing
  force $\force$ ($\rho=0.5$, $A=8$, $L=32$; full lines $U/J=2$,
  dashed lines $U/J=20$; \emph{Inset}: short time evolution.)
  \textbf{(b)} Asymptotic value of $S_t(A)$ (taken at $t=800$) for HCB
  as a function of $V_A$ and compared with the initial bipartite
  entropy $S_0(x_p^+(V_A))$ from both numerics and CFT predictions
  ($\rho=0.5$, $A=200$).}
\end{figure}

After the quench, the condensate is split into two entangled pieces
moving apart: the escaping particles and the self-trapped ones.  This
entanglement between bound states and propagative ones can be
quantified through the bipartite von-Neuman entropy $S_t(A) =
-\text{Tr}\{\varrho_t(A)\ln\varrho_t(A)\}$, where
$\varrho_t(A)\equiv\text{Tr}_{>A}\{|{\Psi(t)}\rangle\langle{\Psi(t)}|\}$.
As seen on Fig.~\ref{fig:entanglement}(a), this entropy essentially
evolves up to an asymptotic value $S_{\infty}(A)$, as expected, which
depends on $N$, $U$ and $\force$.  The effect of the force is
qualitatively inferred from Fig.~\ref{fig:entanglement}(a).  At small
$\force$ when all particles are ejected (see $F=J/4$), after
an initial increase of the entanglement due to the crossing of
position $x=A$ by the many-body wave-packet, the entanglement finally
vanishes when most particles have left region $\cal A$. This
small-$\force$ regime is almost independent on the interaction.
Increasing the force, the asymptotic entanglement first increases,
passing through a maximum when approximately half of the initial
density is ejected, before decreasing at larger forces $\force$,
simply because $N_{\text{esc}}$ falls down. The effect of interactions
on $S_{\infty}(A)$ is two-fold : one is to modify $N_{\text{esc}}$ as
seen on Fig.~\ref{fig:ejected}(d), and the other one is related to the
fact that non-integrability (significant when $U\simeq J$) usually
increases the chaoticity of excited states, leading to higher
entanglement entropies.  A strong enhancement of the entanglement is
thus observed for $U=2J$, by comparison to the behavior close to the
integrable HCB limit (Fig.~\ref{fig:entanglement}(a)). Lastly, in the
inset of Fig.~\ref{fig:entanglement}(a), we show the early evolution
of $S_t(A)$ which is controlled by the rate at which particles are
emitted. This initial emitting rate, as expected, is enhanced by the
repulsion strength $U$ and by the force $F$.

The strong entanglement observed between the two pieces of the
condensate must stem from the initial state in which all particles,
embedded in the same condensate, are naturally entangled. Therefore,
the asymptotic value $S_\infty(A)$ of the bipartite entropy should be
related to the entanglement present initially between the particles
that will leave region $\cal A$ and those that will stay. In the HCB
limit, this idea actually gives a quantitative prediction for
$S_{\infty}(A)$. Indeed, in Fig.~\ref{fig:entanglement}(b) we show
$S_{t=800}(A)\simeq S_{\infty}(A)$, as a function of $V_A$, compared
to the initial equilibrium ground-state entropy $S_0(x_p^+(V_A))$
evaluated at the point $x_p^+(V_A)$ (see Fig.~\ref{fig:ejected}(b))
which marks the leftmost initial position of the ejected
particles. The entropy $S_0$ is computed from exact diagonalization
and also plotted from the (continuous limit) conformal field
prediction~\cite{Calabrese2011}
\begin{equation}
S_{0}(x) = \frac{1}{6}\ln\Big(\frac{4A}{\pi}\sin(\pi\rho)\sin\frac{\pi
  x}{A}\Big) + c'\;, \label{eq:entanglement}
\end{equation} 
with $c'\simeq 0.25$. The good agreement found between these three
curves supports the above picture. The remarkable connection between
the equilibrium and non-equilibrium entanglements found in this setup
could be helpful in the search for measuring many-body
entanglement~(\ref{eq:entanglement}). This connection could be
possibly used to infer equilibrium (local) correlations from
measurements on far apart particle packets.  We also stress that the
setup works for few-particles physics for which the measurement of
entanglement is easier.

\section{Proposals on achieving wave-packet manipulations}

\begin{figure*}[t]
\centering
\includegraphics[width=\textwidth]{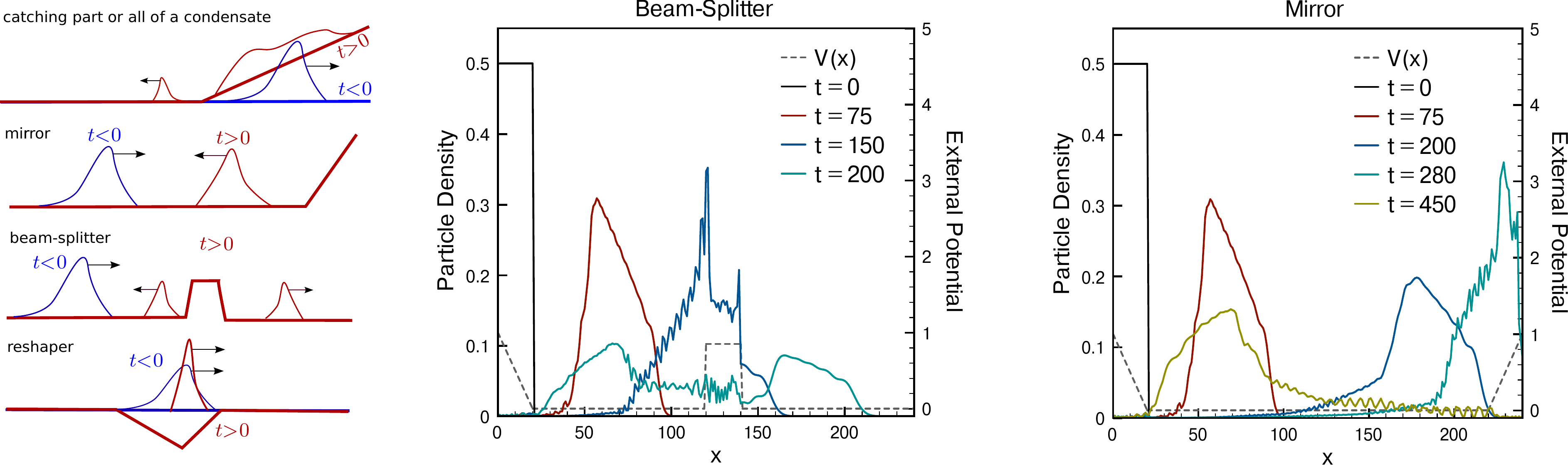}
\caption{(color online) Left : Proposed setups to manipulate a
  travelling condensate on the lattice.  Middle : numerical example of
  the realization of a mirror (the dashed grey line represents the
  $t>0$ potential).  Right : numerical example of the realization of a
  beam-splitter (same color code). \label{fig:spec}}
\end{figure*}

In Fig.~\ref{fig:spec}, we give ideas on how to manipulate travelling
wave-packets generated by the previously discussed ``Galileo ramp''
setup. The arguments are intuitively based on the HCB picture from
energy conservation and demonstrated by numerical calculations. First,
a travelling wave-packet can be catched on a lattice by suddenly
increasing/decreasing the chemical potential (in a ramp shape on the
figure) so that the particles cannot escape the trapping zone. It
requires that the condensate is on the region before changing the
potential (first sketch).  If the potential is raised before the
condensate arrives, then the region acts as a mirror (second sketch).
If the mirror height is finite and a barrier builds up, we obtain a
situation similar to single-particle tunneling process. Then, the
barrier acts as a beam-splitter (third sketch).  Last, applying a
non-uniform potential on a wave-packet can reshape it by slowering the
fastest particles and accelerating the slowest ones (fourth sketch).
This would help fight the natural broadening of the wave-packet.  In
order to show that these intuitive behavior does take place in the
real time evolution of HCB, we give in Fig.~\ref{fig:spec} two
examples of numerical simulations showing the realization of a mirror
and of a beam-splitter. These devices, together with the proposed
source of entangled wave-packets, could help realize interferometry
measurements and their applications -- with the advantage that the two
wave-packets can be spatially well separated, thus being able to
experience different real-space paths.

\section{Damping and thermalization in the closed box geometry}

We see that the main physics is well captured by the integrable HCB
and non-interacting limits.  Still, the model is essentially
non-integrable at finite $U/J$, which has some consequences on the
dynamics.  In order to focus on this aspect, we remove the propagative
band and confine the particles in a box, thus allowing larger sizes
for simulations. In this configuration, BO are damped in the chaotic
regime~\cite{Kolovsky2003} and experiments reported very low damping
nearby the non-interacting point~\cite{Gustavsson2008}. Here, we go
further by computing the damping time and the distance from
thermalization for all $U$.  In particular, we stress that the
integrable nature of the HCB limit shows up in both
quantities. Therefore, we relate in this section the spectral features
of the final Hamiltonian and of the quench distribution to
experimentally accessible observables. In particular, we would like to
see whether the (non)-integrable nature of Hamiltonian has noticeable
consequences on simple observables. In addition to the damping of BO
which was previously discussed in Refs.~\onlinecite{Gustavsson2008,
  Buchleitner2003}, we address the question of the
thermalization~\cite{Dziarmaga2010}.

\begin{figure*}[t]
\centering
\includegraphics[width=0.97\textwidth,clip]{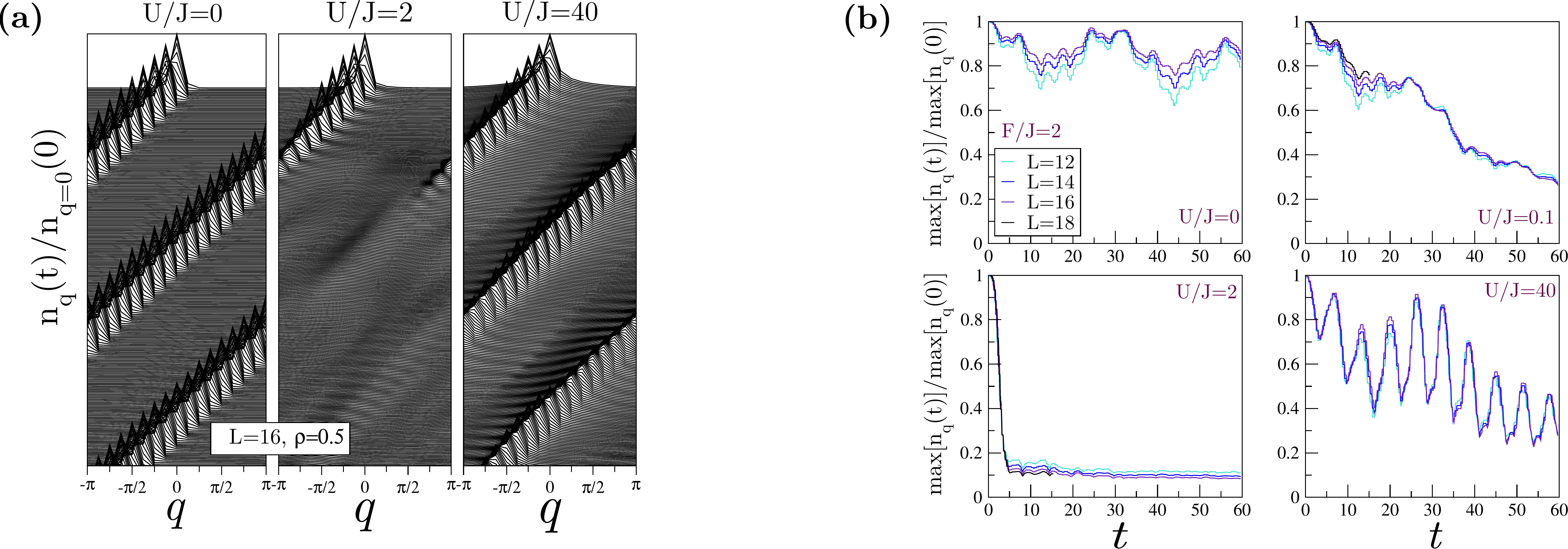}
\caption{(color online) \textbf{(a)} Evolution of the momentum
  distribution $n_q(t)$ as a function of time (vertical shift) for
  different $U/J$ and a fixed force $\force=2J$. The damping seems to
  be clearly related to the non-integrability of the model.
  \textbf{(b)} normalized damping of the momentum distribution $n_q(t)$ for
  different sizes and interactions (the signal local maximum is taken
  over a short time window to remove finite size oscillations
  associated with the $q$-space discretization).}
\label{fig:nk}
\end{figure*}

\subsection{Time evolution after a sudden quench}

The system is prepared in the ground-state $\vert{\psi_0}\rangle$ of a
box without the linear potential which is suddenly turned on at $t=0$.
We write $E_n$ the energies and $\ket{n}$ the eigenvectors of the
final Hamiltonian governing the dynamics. In this basis, the matrix elements
of an observable $O$ are $O_{nm} = \elem{m}{O}{n}$ and the associated
excitation frequencies between $n$ and $m$ are written $\omega_{nm} =
E_n - E_m$. Let $c_{n} = \inter{n}{\psi_0}$ be the coefficients of the
initial state in this basis, and $p_n = \md{c_n}^2$ the corresponding
weights. These weights are the quench distribution, or diagonal
ensemble, and simply provide the averaged contribution of excited
states to the time-evolution. The fidelity $\fidelity$, defined by
\begin{equation*}
\fidelity(t)^2  = \md{A(t)}^2 \;\text{ with }\;
A(t)=\inter{\psi(t)}{\psi_0} = \sum_n p_n e^{-i\omega_{n0}t}\;,
\end{equation*}
plays an important role in the dynamics and contains the information
on the distribution through the Lehmann representation of $A(\omega)$.
Given an Hermitian observable $O$, its real-time behavior and
corresponding derivative read
\begin{align}                                                                                                                                             
\label{eq:phasespace}
O(t) &= \overline{O} + \sum_{n<m} 2\md{c_nc^*_mO_{nm}}\cos(\omega_{nm}t + \phi_{nm}) \\
\label{eq:phasespace2}
\dot{O}(t) &= -\sum_{n<m} 2\md{c_nc^*_mO_{nm}}\omega_{nm}\sin(\omega_{nm}t + \phi_{nm})
\end{align}
with
\begin{equation}
\label{eq:average}
\overline{O} = \sum_{n} p_n O_{nn}
\end{equation}
the time-averaged expectation value (assuming a non-degenerate
spectrum) and $\phi_{nm}$ some phases determined by the initial
conditions and the observable. There are two interesting behaviors in
the time-dependence : the short-time behavior, related to the damping
of the observable towards $\overline{O}$, and the value $\overline{O}$
itself which can be compared to statistical ensemble predictions to
check thermalization. While only the diagonal elements enter the
expression of $\overline{O}$, the off-diagonal ones $O_{nm}$ will play
a role in the damping, together with the associated weights
$\md{c_nc^*_m}$ and frequencies $\omega_{nm}$. These two features
should depend on the integrability of the Hamiltonian through
selection rules, for instance.

\subsection{Observables}

In this section, we define the observables $O$ that are used. The
simplest observable is the local density $ n_j(t) =
\langle{\hat{n}_j}\rangle(t) $. Computing the one-body density-matrix
$g_{jk}(t) = \langle{b^{\dag}_jb_k}\rangle(t)$ allows one to access
two important observables. The first one is the momentum distribution
$n_q(t) = \langle{\hat{n}_q}\rangle(t)$, with
\begin{equation}
n_q(t) = \frac 1 L \sum_{jk} e^{iq(j-k)} g_{jk}(t)\, 
\end{equation}
which is measured using time of flight techniques. A peak in $n_{q}$
usually measures the coherence of the condensate at the corresponding
wave-vector. The second one is the condensate fraction $f_0(t)$,
defined as the largest eigenvalue of the one-body density-matrix. It
matches $N$ when all bosons share the same state, and is of order
$\sqrt{N}$ in the HCB regime.  Even so it is not experimentally
observable, the fidelity $\fidelity(t)$, defined above, quantifies the
distance from the initial state and is as well sensitive to the
chaotic features of the Hamiltonian.

\subsection{Damping of observables}

The spectral features and chaoticity of the Hamiltonian are provided
in Appendix~\ref{app:spectral}.  When one goes to the large-$U$ or
large-$F$ limits, the frequencies spectrum $\omega_{nm}$ is such that
they are all very close to the same level spacing ($U$ or $F$), i.e. a
nearly equally spaced spectrum. Of course, as judged by the poissonian
nature of the spectrum at finite $J$, a dense level-spacing
distribution is still present, generically leading to damping. In this
regime, we expect some observables to show strong oscillatory behavior
at the main frequency and its harmonics. This is a situation
qualitatively related to the classical motion in closed orbits. In the
quantum version, this would show up in the pseudo phase-space
$(O,\dot{O})$ (from Eqs.~\eqref{eq:phasespace}-\eqref{eq:phasespace2})
as more or less complex orbits depending on the number and weights of
the harmonics involved. In the nearly equal level spacing spectrum
situation, the fidelity would obviously display similar resonances (or
revivals). In the case of non-integrable Hamiltonian, although level
repulsion tends to rigidify the spectrum, we expect that the weights
and the frequencies are spread and of the same magnitude so that the
damping of observables usually occurs on shorter times. In the pseudo
phase-space $(O,\dot{O})$, damping translates into a spiral structure
of the orbit. We do observe this qualitative effect, particularly on the
local density and its associated particle current (data not shown).

The first interesting quantity to look at is the momentum distribution
v.s. time because it captures well the Bloch oscillations.  We recall
that, at the single-particle level, the momentum is shifted with time
according to $q(t)=q(0)+\force t$.  In Fig.~\ref{fig:nk}, we recover
that $n_q(t)$ is globally shifted with time for all $U/J$, with a
period close to the Bloch period $\tau_B$. Increasing $U$ further does
not change much this period (hardly seeable in the numerics) and the
main effect of interactions is to destroy the main peak signaling the
bosonic coherence. Indeed, for $U=2J$, we observe a strong damping of
the coherence after short times. Interestingly, increasing $U$ brings
us close to the HCB limit and the damping time increases
again. Finally, for very large $U$, the central peak shape is
preserved for very long times.

\begin{figure}[t]
\centering
\includegraphics[height=0.65\columnwidth,clip]{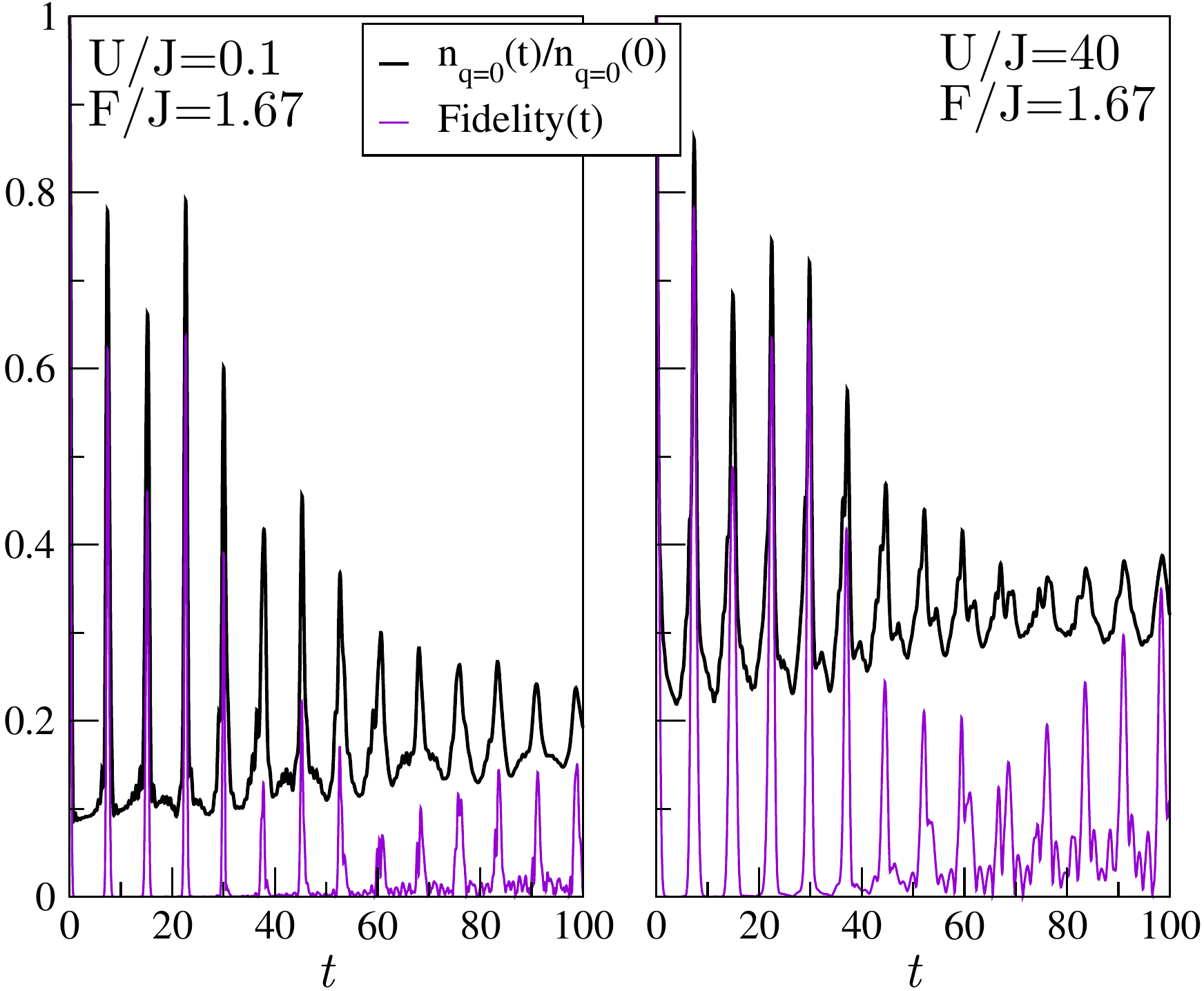}
\caption{(color online). Comparison between $n_{q=0}(t)$ and the fidelity in the
  integrable regimes, showing the revivals.}
\label{fig:nk2}
\end{figure}

Another interesting point with the momentum distribution within this
setup is that the zero-momentum evolution is qualitatively related to
the fidelity of the system in the near-integrable regimes, both
signaling strong revivals of the initial wave-function (see
Fig.~\ref{fig:nk2}). In fact, we know that the momentum distribution
gets back to its initial position in $q$-space after a period
$\tau_B$. More than the momentum distribution, the whole many-body
wave-function actually comes back close to the initial state, yielding
strong revivals in the fidelity. We observe that the peaks in the
fidelity and in $n_{q=0}$ are nicely correlated signals in these two limits.

\begin{figure}[t]
\centering
\includegraphics[width=0.7\columnwidth,clip]{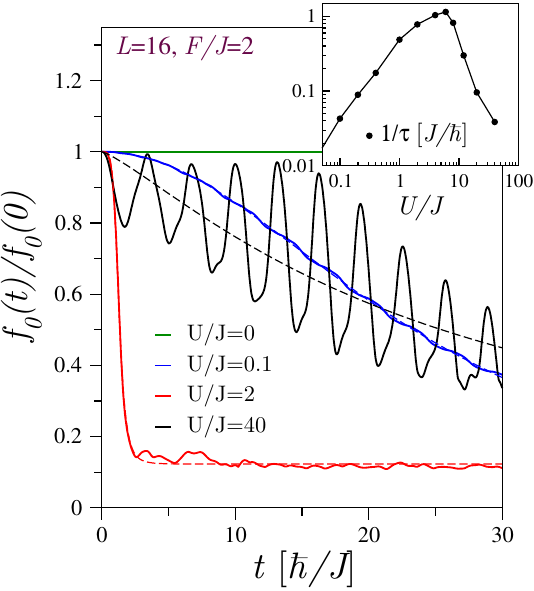}
\caption{(color online). Typical evolution of the condensate
  fraction $f_0(t)$ with time for various interactions and fixed ramp
  $\force=2J$. \emph{Inset}: inverse damping time $\tau$ v.s. $U/J$.
  \label{fig:dampingtime}}
\end{figure}

Fig.~\ref{fig:dampingtime} shows the damping of the condensate
fraction $f_0$ for increasing $U/J$ and fixed force.  The behavior is
very similar to the momentum distribution, but easier to fit to
extract a typical damping time $\tau$. We extract the typical damping
time from the normalized condensate fraction time-evolution by fitting
the curves using the function $f(t)=a+(1-a)/(1+(t/\tau)^b)$. $a$ is
the asymptotic value of the fonction, and $\tau$ is such that the
curve has decayed of half its distance from 1 to $a$. The exponent $b$
is found to vary between 1 and 5 in the fits but its value does not
significantly change $\tau$ which is the quantity of interest. Good
fits are obtained at small $U$ while oscillations spoil the fit at
large $U$, which nevertheless provides a reasonable estimate of the
damping time. The main plot and the inset supports that the $\tau$s
become much longer near the two integrable points.

\subsection{Thermalization}
\label{sec:thermalization}

In addition to the short-time behavior, the difference between
integrable and non-integrable regimes can show up in the long-time
average value $\overline{O}$. We see that both the features of the
diagonal distribution $p_n$ and the behavior of the observables with
energy intervene in Eq.~\eqref{eq:average}. If one wants to study
thermodynamics only, the energy distribution $p_n$ is the only useful
quantity. In the case of a generic distribution, the equivalence of
ensemble should be sufficient to provide the same thermodynamics as
usual thermal ensembles~\cite{Roux2010a}. If one wants to compare
time-averaged observables $\overline{O}$ to those obtained from
thermal ensembles, the hypothesis that $O_{nn}\sim O(E)$ hardly varies
within the energy shell given by the typical energy fluctuations
naturally leads to the identification with $\overline{O}$ and,
therefore, to thermalized observables in this sense. Justification of
this idea (sometimes dubbed as the Eigenstate thermalization
hypothesis, or ETH) has been proposed from quantum chaos
arguments~\cite{Peres1984, Deutsch1991} and studied
numerically~\cite{Feingold1984, Rigol2008}. Some extensions and more
details prescription of this statement have been discussed more
recently~\cite{Biroli2010}, stressing the importance of the
correlations between $p_n$ and $O_{nn}$. In this context,
integrability or the proximity to integrable points on finite systems
has been shown to lead, most of the time, to non-thermalized
observables~\cite{Rigol2006, Rigol2008, Santos2010, Rigol2009,
  Iucci2009, Roux2010, Santos2010a, Biroli2010}.  The particular
situation of an interaction quench in the Bose-Hubbard model which
possesses integrable limits, has been discussed in depth in
Refs.~\onlinecite{Kollath2007, Roux2009, Roux2010, Biroli2010}. Of
course, the above approaches suffer from limitations in their
applicability. For instance it is expected to work best at high
energies where quantum chaos describes well the states. Yet, this
corresponds to high-temperature where the physics is in general not so
interesting. Another issue comes when taking the thermodynamic limit
$N \rightarrow \infty$ after or before time-averaging.

\begin{figure*}[t]
\centering
\includegraphics[width=\textwidth,clip]{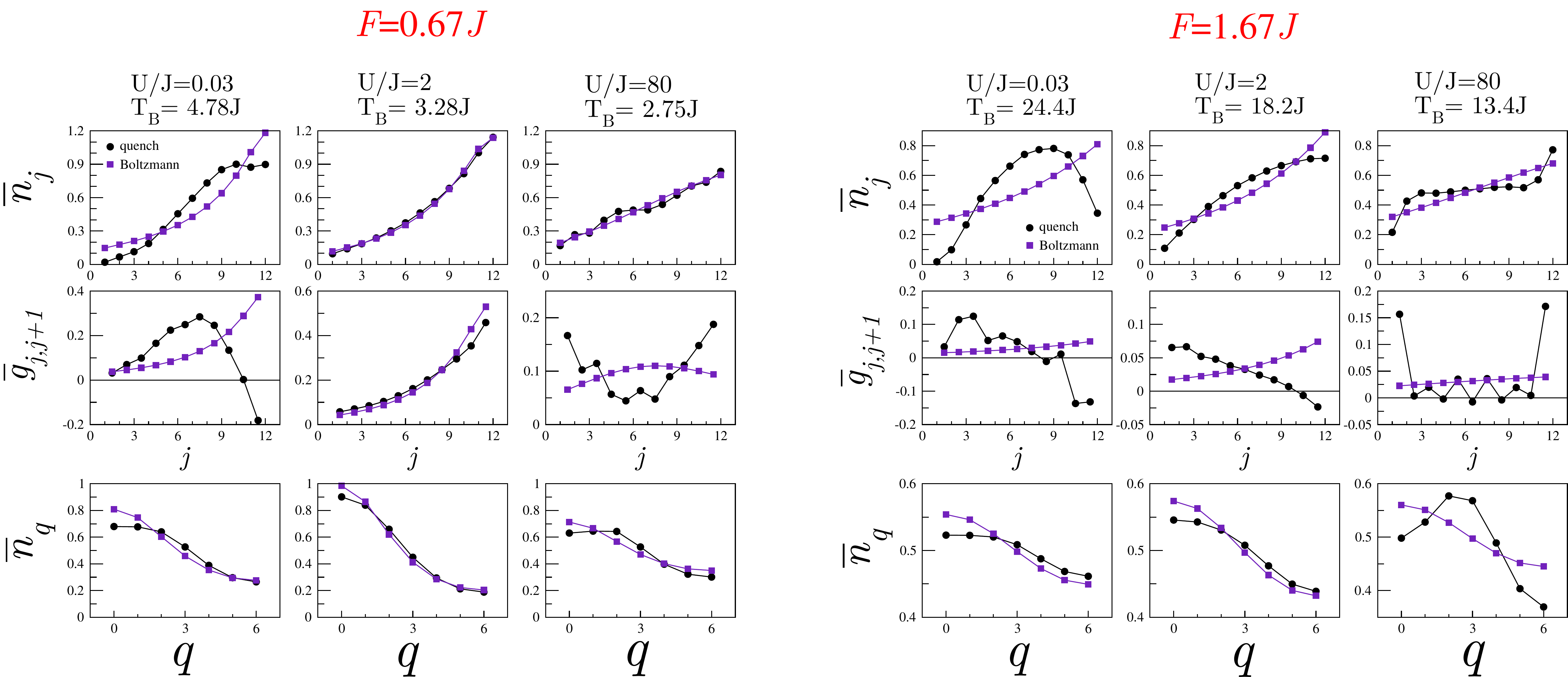}
\caption{(color online). Comparison between time-averaged profiles
  obtained after the quench and thermalized profiles with the same
  mean-energy (in a closed box). \emph{Left}: for a force $\force$
  large enough to put enough energy in the system, but not too large
  to destroy non-integrability, the gas looks thermalized in the
  non-integrable regime $U/J=2.0$, while clear deviations from
  thermalization are observed close to the integrable limits
  $U=0,\infty$. \emph{Right}: When the force is larger, this
  finite-size system ($L=12$) looks too integrable to observe
  approximately thermalized behaviors. In both cases, $T_B$ indicates
  the temperature obtained from the mean-energy using the canonical
  ensemble.}
\label{fig:profiles}
\end{figure*}

We now turn to the calculation of the time-averaged expectation
$\overline{O}$ of an observable compared to its thermal prediction
$O^{\text{th}}$. $\overline{O}$ is directly obtained from
\eqref{eq:average} from full diagonalization of the Hamiltonian on
small systems (we use $L=12$ and $N=6$) and from computing the quench
distribution $p_n$ exactly. $O^{\text{th}}$ is obtained using
calculation in the canonical ensemble. A microcanonical description is
also possible but it is in practice less transparent as one has to
tune to energy window. Usually, the outcome does not depend much on
the choice of the distribution at sufficiently high energies
(temperatures) and looking at simple observables; Besides, the
equivalence of ensemble suggests that both ensemble should in
principle yield the same results in the thermodynamical limit,
i.e. for large enough systems. The temperature $T_B$ of the Boltzmann
distribution is directly obtained by demanding that the mean-energy
should be the same as the mean-energy of the system after the quench
$\langle{E}\rangle = \sum_n p_n E_n$. Then, the Boltzmann weights read
$p_n = e^{-E_n/k_BT}/Z$ with $Z$ the partition function and we plug
them in \eqref{eq:average} to get $O^{\text{th}}$.

The goal is now to see whether deviations from thermalization are
related to the integrability of the model studied previously. In
Fig.~\ref{fig:profiles}, we show the profiles for the three simplest
observables for quenches with moderate forces $\force$ and three
typical values of $U/J$ respectively close to the non-interacting
limit, deep in the non-integrable region and close to the HCB
limit. One must not choose a force which is too small, otherwise
little energy is put in the system, the temperature would be too small
and the discussion possibly spoiled by finite-size
effects~\cite{Roux2010}. If the force is too large, we know from
Fig.~\ref{fig:integrability} that the system looks integrable for all
$U/J$. We find that $F/J \simeq 0.5,1$ is a reasonable compromise to
use the quench protocol to probe the relation between thermalization
and integrability in this setup. Indeed, in Fig.~\ref{fig:profiles},
we observe that for $F/J=0.67$, all profiles agree well with the
thermal predictions for $U/J = 2$, deep in the non-integrable
regime. Close to the integrable points, deviations are found but the
effect depends on the chosen observable. We find that the local
kinetic energy $g_{j,j+1}$ is the most suitable observable to probe
the effect of the integrability (something also discussed in
Ref.~\onlinecite{Santos2010a}). For $F/J=1.67$, it is much harder to
reach thermalization, probably due to the near integrable features of
the Hamiltonian for this small system.

\begin{figure}[t]
\centering
\includegraphics[width=0.65\columnwidth,clip]{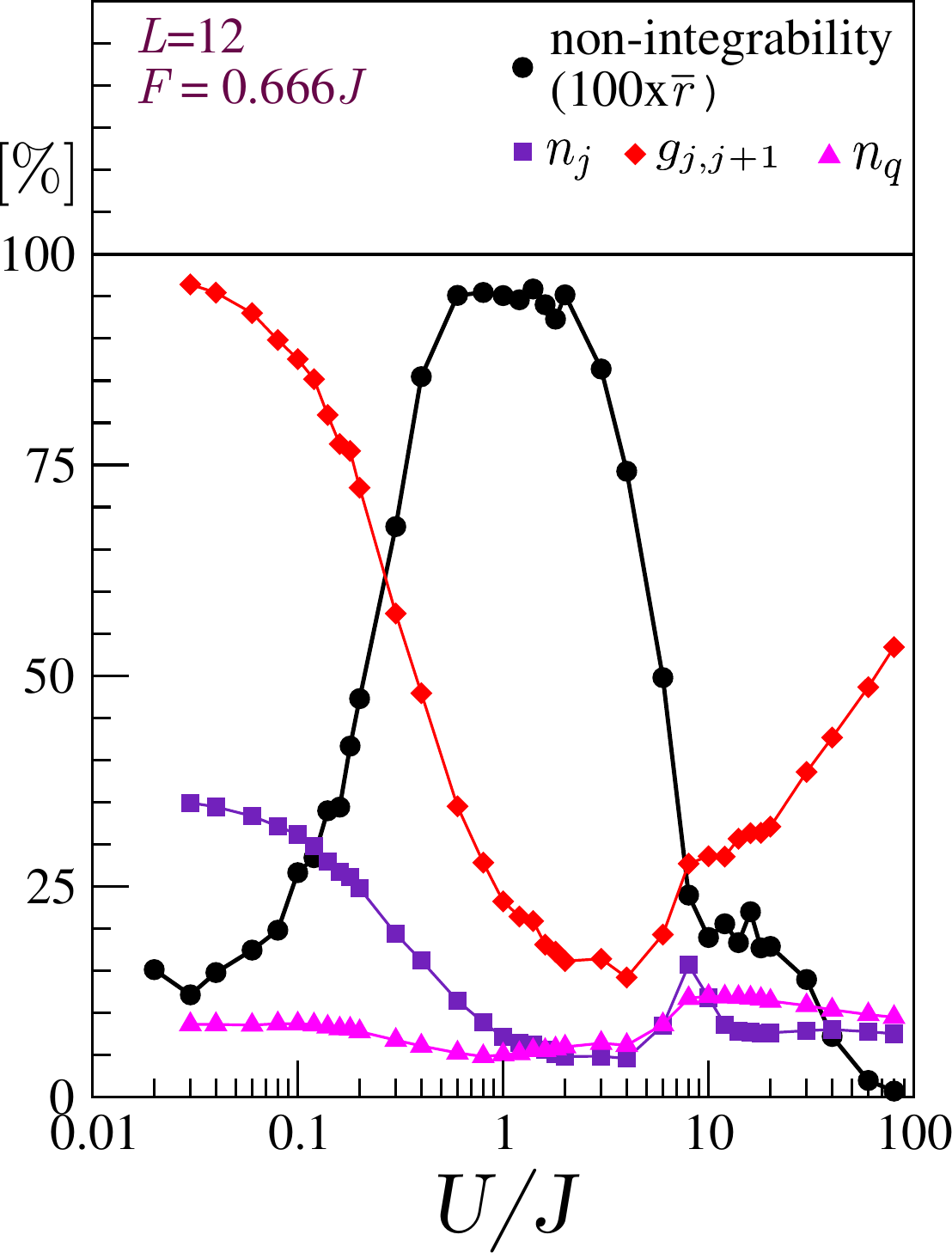}
\caption{(color online). Relative distance (in \%) between the time-average and the
  thermal expectations of an observable v.s. $U/J$ (density $n_j$,
  kinetic energy $g_{j,j+1}$, momentum distribution $n_q$). $\bar{r}$
  measures the non-integrability of the model.
  \label{fig:thermalization}}
\end{figure}

In order to quantify the deviation from the thermal prediction, we
average over the sites $j$ (or momentum $q$ for $n_q$) the relative
distance from thermal prediction and dub it:
\begin{equation}
\delta O = \frac 1 L \sum_j^L \left\vert{\frac{\overline{O}_j-O_j^{\text{th}}}{O_j^{\text{th}}}}\right\vert\;.
\end{equation}
This is the quantity plotted in percentage in
Fig.~\ref{fig:thermalization} as a function of $U/J$.  We
qualitatively expect that non-thermalized regimes occur close to
integrable points~\cite{Rigol2008,Rigol2009,Roux2010,Santos2010a}.
This is well observed in Fig.~\ref{fig:thermalization} for which the
integrability measure, denoted by $\bar{r}$ (see
Appendix~\ref{app:spectral} for definitions), is manifestly correlated
to the non-thermalization measure. The advantage of this setup with
respect to interaction quenches~\cite{Kollath2007, Roux2009, Roux2010,
  Biroli2010} is that the ratio $U/J$ is kept fixed. In other words,
it decouples the quenching parameter from the parameter which mainly
controls integrability.

\section{Conclusion} 

We introduced and characterized a remarkable setup, the ``quantum
Galileo ramp'', which can be used to create entangled many-body
wave-packets. We further provide simple views on how to manipulate the
condensate afterwards. An important result is that the asymptotic
non-equilibrium entanglement entropy is quantitatively related to the
equilibrium initial one in the HCB limit. In addition, non-trivial
Josephson-like oscillations are found and we show the setup is well
suited to study the role of integrability on damping and
thermalization. We underline that results on HCB concerning the local
density are also valid for free fermions. Beyond the field of cold
atoms, the setup could be developed for polariton
condensates~\cite{Kasprzak2006}.

\subsection*{Acknowledgements}

This work is supported by ANR-09-BLAN-0098-01. M.~C. benefited from
the International Graduate College on Statistical Physics and Complex
Systems between the universities of Nancy and Leipzig.

\appendix

\section{Neglecting the tunneling toward the upper Bloch band}
\label{app:tunneling}

Experimentally, tunneling to upper Bloch band could occur if the band
is to close in energy to the states of the trapped particles. Indeed,
under the constant force $F$, Bloch oscillations may be ruled out by
Landau-Zener interband transitions (and particles may possibly escape
into the continuum if the upper gaps are too small). The rate at which
this escape toward the upper band is expected is roughly given by the
Landau-Zener tunneling formula $e^{-c \delta^2/F}$
\cite{BenDahan1996,Peik1997} where $\delta$ is the interband energy
gap and $c$ is a constant depending in particular on the recoil energy
($E_R=\Delta$ here). To avoid the escape of particles toward the upper
band one needs accordingly to apply a sufficiently low force such that
$F\ll c \delta^2$. On the other hand, one needs a sufficiently high
force in order to scan the first Brillouin zone within a reasonable
small time (from an experimental point of view) \cite{Peik1997}.

\section{Spectral features and chaos in the box geometry (without a propagative band)}
\label{app:spectral}

\subsection{Hamiltonian}

We study using Lanczos and full diagonalization techniques the
Bose-Hubbard Hamiltonian in box of size $L$ and subjected to a linear
potential $V_j = V_A - \force j$, where $\force$ is the slope (or
force):
\begin{equation}
\label{eq:ham}
\mathcal{H} = -J \sum_j [b^{\dag}_{j+1} b_j + \text{h.c.}]
+ \frac U 2 \sum_j n_j(n_j-1)+ \sum_j V_j n_j 
\end{equation}
with $V_j$ the external potential. If the onsite boson cutoff $M=1$,
we have the integrable XX model while if $M=N$ we recover the
Bose-Hubbard model. In the following, all data are for a density
$\rho=N/L=1/2$ (superfluid regime) and full diagonalization are done
with $L=12$ (Hilbert space size of about 12000 states), and Lanczos
calculations up to $L=18$.

This model has been investigated in a similar context, analyzing the
effect of interactions on BO and its regular/chaotics regimes. The
main known results are that a new period, in addition to $\tau_B$,
emerges in the BO for finite $U$ and in the strong force limit, which
reads $2\pi/U$~\cite{Kolovsky2003}. Chaotic motion is found for small
$\force$ and in the presence of interactions, leading to a damping of
the BO~\cite{Buchleitner2003}. A strong reduction of damping has been
observed experimentally~\cite{Gustavsson2008} when $U\rightarrow
0$. The Mott regime $U\gg J$ of the BO has also been briefly
investigated in Ref.~\onlinecite{Kolovsky2004}. Level statistics and
the chaotic nature of the spectrum has been partially analyzed in
Ref.~\onlinecite{Buchleitner2003,Kolovsky2003a}. Approximate methods such as
discrete non-linear Schr\"odinger equation and mean-field theory have
been also used to study the dynamics of this
model~\cite{Kolovsky2004,Kolovsky2009,Kolovsky2010}. We lastly mention
that the hard-core boson dynamics in a tilted bichromatic lattice
(which possesses a gap in the single-particle dispersion relation) has
been investigated in Ref.~\onlinecite{Cai2011}.

\subsection{Density of states}

\begin{figure}[t]
\centering
\includegraphics[width=0.75\columnwidth,clip]{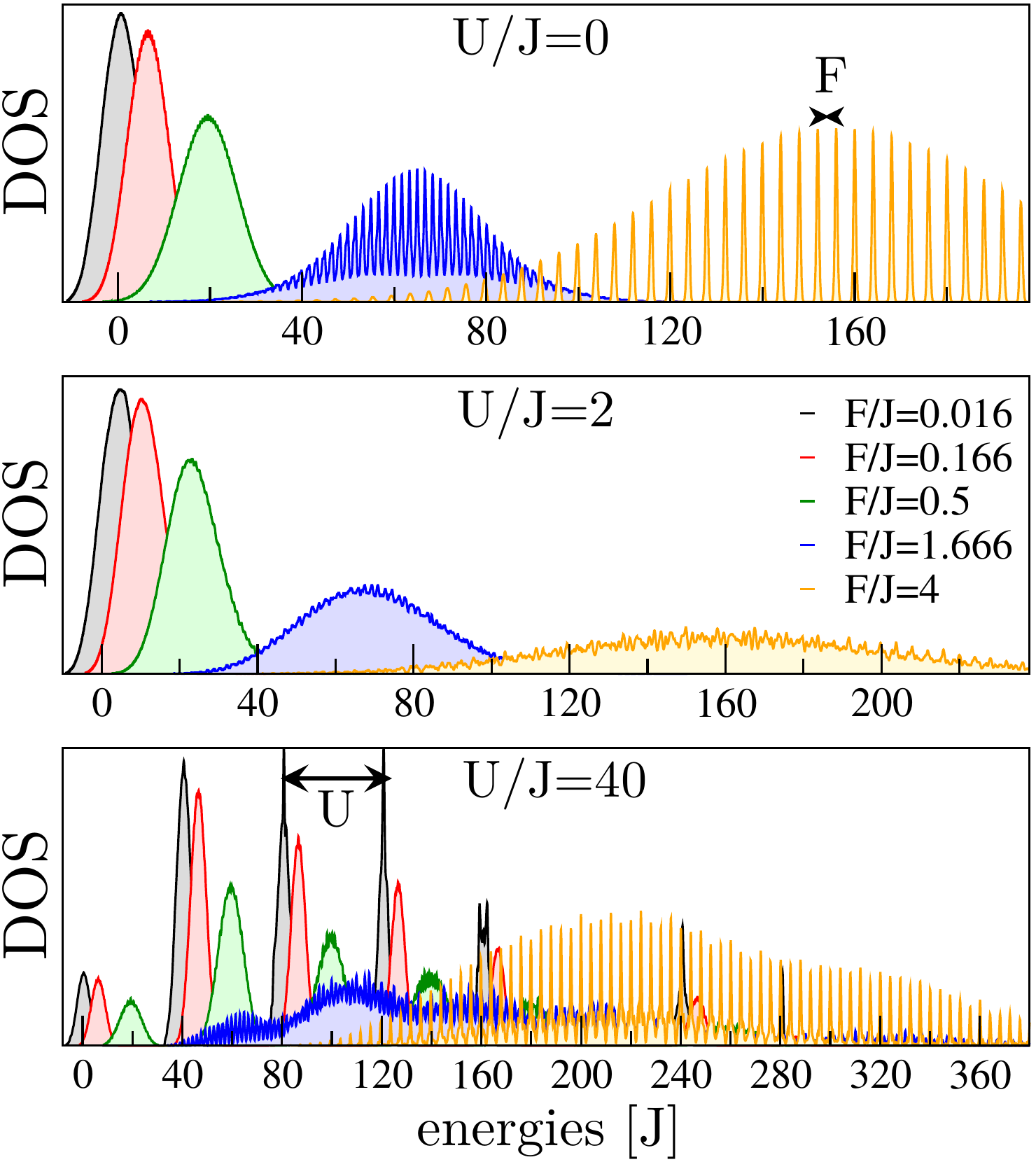}
\caption{(color online). Evolution of the density of states of the many-body
  Hamiltonian with slope $\force$ for different $U$. We observe main
  bands separated by $U$ at large $U$ or by $\force$ at large
  $\force$.}
\label{fig:dos}
\end{figure}

We first discuss the nature of the many-body density of states (DOS)
of the Hamiltonian as a function of the two parameters $U/J$ and
$F/J$. There are three energy scales in the problem : the hopping $J$
which sets the bandwith of the single-particle energies, the local
interaction $U$ and the external potential ramp $\force$. These
typical energies are visible in the many-body DOS of the spectrum as
seen in Fig.~\ref{fig:dos}. When only $J$ is present, the DOS has a
gaussian shape centered around zero energy. In the large-$F$ limit, we
observe equally-spaced peaks separated by the energy $F$. This
structure is reminiscent of the single-particle Wannier-Stark ladder
spectrum. It is as well trivially understood at the many-body level
when $F\gg J$ since $F$ corresponds to the potential energy cost when
a particle jumps from one site to its neighbor. Similarly, when $F=0$
and $U\gg J$, elementary processes corresponding to changing the
onsite number of particles yield an equally-spaced spectrum of energy
$U$ (Mott lobes). Notice that $F$ rapidly kills these lobes as we see
on Fig.~\ref{fig:dos} that $F/J\gtrsim 1$ is sufficient to destroy the
lobes in the DOS.

\subsection{Signatures of integrability and chaoticity in the spectrum}

In order to study the non-integrable nature of the Hamiltonian
\eqref{eq:ham} which governs the time-evolution, we use level and
wave-function statistics (not shown). However, we see that the usual
unfolding procedure for the spectrum will be plagued in the large
$\force$ or large $U$ regimes, due to the peak structure of the
DOS. Following Refs.~\onlinecite{Oganesyan2007,Kollath2010}, we use
the statistics of the ratio of consecutive level spacings $r_n =
\min(\delta_n,\delta_{n-1}) / \max(\delta_n,\delta_{n-1})$ where
$\delta_n = E_{n+1}-E_{n}$ is the level spacing and $E_n$ are the
energies. Probability distributions $P(r)$ are then compared to
poissonian and gaussian orthogonal ensemble (GOE) statistics. 

\begin{figure}[t]
\centering
\includegraphics[width=0.85\columnwidth,clip]{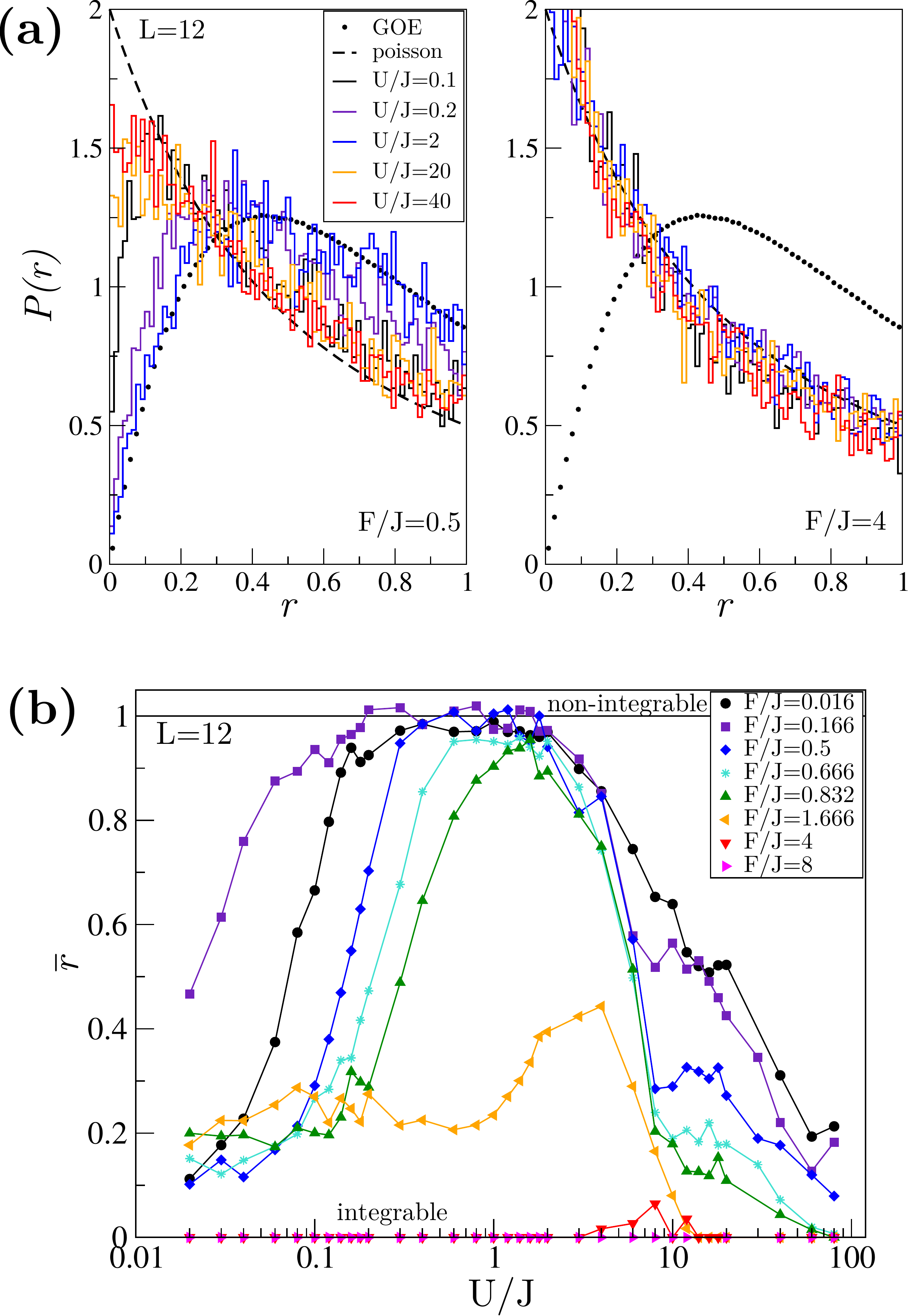}
\caption{(color online). \textbf{(a)} Probability distributions of the
  mean adjacent level spacing ratio $P(r)$ as a function of $U/J$ for
  two different slopes $\force$. \textbf{(b)} Evolution of the
  normalized mean adjacent level spacing ratio $\langle{r}\rangle$ as
  a function of $U/J$ for different slope $\force$.}
\label{fig:integrability}
\end{figure}

An example for the model under study is given in
Fig.~\ref{fig:integrability} in which we observe both regular
(Poisson) and chaotic (GOE) distributions, depending on the
parameters. In order to extract a simple number to be plotted against
the parameters, we use the normalized mean ratio defined by $\bar{r} =
(\langle{r}\rangle-\langle{r}\rangle_{\text{Poisson}}) /
(\langle{r}\rangle_{\text{GOE}}-\langle{r}\rangle_{\text{Poisson}})$,
which is 1 in the non-integrable regime and 0 in the integrable
regime.  When $\force\neq 0$, the reflection symmetry of the box is
lost and the Hamiltonian \eqref{eq:ham} has no spatial symmetries
which could plague the level statistics if not taken into account.  As
discussed in Ref.~\onlinecite{Kollath2010} in the $\force=0$ case,
only the $U=0$ and $J=0$ limits are integrable, Bethe-ansatz
predicting the Hamiltonian is non-integrable as soon as $U,J \neq
0$. Still, because of finite size effects, small and large $U/J$
regions look almost integrable on a finite cluster. The thermodynamic
limit is particularly hard to investigate numerically, although the
trend is compatible with the Bethe-ansatz predicting the Hamiltonian
is non-integrable as soon as $U,J \neq 0$. Still, because of finite
size effects, small and large $U/J$ regions look almost integrable on
a finite cluster. The thermodynamic limit is particularly hard to
investigate numerically, although the trend is compatible with the
Bethe-ansatz perspective~\cite{Kollath2010}. We then
expect~\cite{Buchleitner2003, Kolovsky2003a} the large $U$ and large
$\force$ regime to be close to the integrable classical limit ($J=0$),
at least on a finite chain. The results for different $\force$ as a
function of $U/J$ are displayed on Fig.~\ref{fig:integrability} and we
see that for $\force=4J$, the distribution looks always poissonian. In
order to observe non-integrable effects, we must not choose $\force$
too large ($\force\lesssim 1$ for this $\rho=1/2$ and $L=12$). We also
notice from this figure that the optimal range of $U$ to be in the
non-integrable regime depends on $\force$.

\end{document}